\documentclass[aps,prl,twocolumn,showpacs]{revtex4-1}
\usepackage{graphicx} 
\usepackage{amssymb}
\usepackage{float}
\begin{document}
\title
{\bf Radial Line Nodes in Weakly Anisotropic Noncentrosymmetric Superconductors}
\author{Mehmet G\"{u}nay$^{\bf (1)}$ and T. Hakio\u{g}lu$^{\bf (2)}$}
\affiliation{ 
${\bf (1)}$ {Department of Physics, Bilkent University, 06800 Ankara, Turkey}
\break
${\bf (2)}$ {Consortium of Quantum Technologies in Energy (Q-TECH), Energy Institute, \.{I}stanbul Technical University, 34469, \.{I}stanbul, Turkey}
}
\begin{abstract}
Noncentrosymmetric superconductors (NCSs) without inversion symmetry (IS) have a doublet of mixed parity order parameters (OPs) which can have  nodes. In addition to the angular line nodes (ALNs) existing under strong anisotropy, radial line nodes (RLNs) exist in weakly anisotropic NCSs due to the radial momentum dependence of the interactions and the broken IS. We study the topology, the number and the positions of RLNs which can be controlled by the chemical potential and the degree of IS breaking. The RLNs exhibit a low temperature behaviour intermediate between exponential suppression and the integer powerlaw. For this reason they are difficult to detect and may be inadvertently missed in a number of experiments. We show that Andreev conductance experiments can efficiently distinguish RLNs in the energy gap from the other fundamental nodes. 
\end{abstract}
\maketitle

Superconducting symmetries beyond the conventional BCS spin singlet pairing are known since 1960s. Distinct examples are $^3He$ \cite{He3}, heavy fermion \cite{Heavy_Fermion}and high $T_c$ \cite{High_Tc}superconductors all of which are strongly correlated systems. However, and unlike BCS, in this case there is not a single universal mechanism, the source of which can be the momentum dependent interactions due to the strong electronic correlations in the presence of anomalies as broken spin-degeneracy, broken IS and the spin-orbit coupling (SOC)\cite{TH}. The time reversal symmetry (TRS) can be externally or spontaneously broken\cite{TRSB}. Particularly, NCS, as the inversion symmetry broken subset, embody a rich variety of crystal symmetries and interactions that can reveal spin dependent OPs\cite{OPs} with mixed parity. 

The smoking gun of the unconventional pairing in an NCS is in the nodes related to its OPs\cite{evidence1,evidence2}. There, the OPs have mixed parities in the presence of broken IS and finite SOC with coexisting even singlet ($\psi_{\bf k}$) and the odd triplet (${\bf d}_{\bf k}$). The OPs can most simply be represented as the sum $\psi_{\bf k}=\sum_m \, \psi^{(m)}_{\bf k}$ and ${\bf d}_{\bf k}=\sum_m \, {\bf d}^{(m)}_{\bf k}$ with $\psi^{(m)}_{\bf k}={\cal Y}_m({\hat {\bf k}})\psi^{(m)}_k$, ${\bf d}^{(m)}_{\bf k}={\cal Y}_m({\hat {\bf k}}){\bf D}^{(m)}_{\bf k}$ and ${\hat {\bf k}}={\bf k}/k$ and ${\cal Y}_m$'s, with $m=0,\pm 2, \pm 4, \dots$, are angular basis functions describing the anisotropy\cite{Sigrist_Ueda} and $\psi^{(m)}_k$ and ${\bf D}^{(m)}_{\bf k}=({\bf D}^{(m)}_{x,\bf k},{\bf D}^{(m)}_{y,\bf k},{\bf D}^{(m)}_{z,\bf k})={\bf k}{\cal D}^{(m)}_k$ are  pairing functions for each angular momentum component $m$. In two dimensional bulk NCSs, which we consider here, the nodes are realized vertically, i.e. ${\bf k}=(k_x,k_y,k_z)$ with $k_z=0$ and classified as a) discrete set of ${\bf k}$ points, i.e. point nodes (PNs), and b) line nodes. Under manifested TRS, the excitation spectra are Kramers degenerate $E_{\bf k}^\lambda=E_{-\bf k}^{\lambda^\prime}$ with $\lambda, \lambda^\prime$ describing different branches split by the broken IS. Furthermore, excitation spectra are symmetric around all TRS-invariant points such as ${\bf k}=0$ and ${\bf k}=\{(\pm \pi,0), (\pm 0,\pi)\}$. PNs can occur at these TRS-invariant points or, in the case of Weyl points they may occur at arbitrary values in TRS  pairs\cite{Weyl}. 

In addition to the PNs in two dimensional NCSs, and if $m\ne 0$ components of the OPs are sufficiently strong, ALNs can be formed in certain momentum directions where the OPs vanish. For instance, in the tetragonal $C_{4v}$ which ALNs can be formed at $k_x=\pm k_y$ or $(k_x=0, k_y=0)$.  
In low temperatures, ALNs are known to cause quadratic scaling exponents in the $C_V$ and this has been confirmed in a number of highly anisotropic NCSs\cite{evidence1,evidence2} among which are the celebrated TRS preserving $CePt_3Si$ and the TRS breaking $Sr_2RuO_4$. They also cause power law behavior in London penetration depth, heat conductivity and ultrasound attenuation with integer exponents. However, a large number of experiments also exist where ALNs are insufficient to explain the data.\cite{Sigrist_Ueda} 

The other extreme in NCSs is the weak or no anisotropy in which case no ALNs are present and the $m=0$ component dominates with ${\cal Y}_0({\hat {\bf k}})\simeq 1$ and negligible ${\cal Y}_{m\ne 0}({\hat {\bf k}})$. Then $C_{\infty v}$ is a good symmetry where $\vert\psi_{\bf k}\vert \simeq \psi_{k}$ and $\vert{\bf d}_{\bf k}\vert \simeq F_{k}$ are invariant under rotations. In this case however, nodes are at fixed radial positions in the mixed state, i.e. radial line nodes (RLNs). They arise due to the nontrivial $k=\vert {\bf k}\vert$-dependence of the OPs\cite{TH_MG}. The RLNs can be present in $\psi_k$ and/or $F_k$ or, in the case of strong IS breaking, in the mixed state energy gaps $\vert \tilde{\Delta}_{k}^{\pm}\vert =\vert \psi_{k} \mp \gamma_{k} F_{k}\vert$. Here $\pm$ is the band splitting due to the broken IS and $\gamma_k$ is a function of $k$ which can only take the values $\pm 1$ (see below). From now on, we will only consider the $m=0$ case. If $\psi_{k}=\pm F_{k}\ne 0$ at a certain position $k=k_n$, RLNs occur in one or both of the $\tilde{\Delta}_{k}^{\pm}$. A node in $\tilde{\Delta}_{k}^{\pm}$ is a consequence of a high triplet/singlet ratio $R_{k}=\vert {F}_{k}/ \psi_{k}\vert=1$ at $k=k_n$ which indicates a strong IS breaking. For directly identifying RLNs a precise handling of the $k$ dependence in the interactions and the OPs is required\cite{TH_MG}. Between the strong and weak anisotropic limits, angular and the radial line nodes can coexist. Hence, distinct experimental signatures that can separate these two different types of line nodes are highly valuable for deeper understanding of the NCSs, and to our knowledge, this has not been pursued yet. This letter focuses on the topological and the thermodynamic properties of RLNs and their distinctions from the angular ones.          

In general, whether point or line, the nodes can be present in three distinct levels: a) the OPs $\psi_{\bf k}, {\bf d}_{\bf k}$, b) the $\tilde{\Delta}_{\bf k}^{\pm}$ and c) the energy spectra $E_{\bf k}^\pm$. Examining the singlet and the triplet nodes in (a) separately is experimentally very difficult. On the other hand, nodes in the second class (b) can be most accurately identified by the ARPES\cite{ARPES}, Andreev conductance (AR)\cite{Andreev_cond} or other ingenious measurements\cite{Kurter}. In the third class (c), the nodes in $E_{\bf k}^\pm$ can be searched for in measurements that are sensitive to the energy density of states (DOS) $\rho(E)$. Power laws in temperature with integer exponents in the specific heat and the superfluid density  were suggested for point and line nodes\cite{Bauer_Sigrist} and observed in compounds with strong planar anisotropy\cite{nodal_thermodynamic,strong_anisotropy}. On the other hand, some data on weakly anisotropic compounds\cite{weak_anisotropy} can be less confidently fitted to exponentially suppressed BCS form but also to a sum of different power laws. The existing theories can therefore become inadequate in explaining a number of data.\cite{inadequate} It is also important to remember that, a node in $\tilde{\Delta}_{k}^{\pm}$ does not imply a certain nodal structure in $\psi_{k}, F_{k}$ and vice versa. However, the rare circumstance of a node in $E_{\bf k}^{\pm}$ does imply a node in $\tilde{\Delta}_{\bf k}^{\pm}$ and not vice versa.   

Distinct features that make RLNs important in this picture become clear in their topological analysis. Considering a manifested TRS requires that $\psi_k\,d_{z,{\bf k}}=0$, hence $d_{z,{\bf k}}=0$ \cite{TH_MG}. With a proper choice of the basis, the mixed state Hamiltonian can be block-diagonalized as $H={\bf h}_{\bf k}^{\pm}.{\bf \sigma}$  where $\pm$ are the SOC branches, and ${\bf h}_{\bf k}^{\pm}=(h_x,h_y,h_z)=(\tilde{\Delta}_k^{\pm} \cos\phi_{\bf k},-\tilde{\Delta}_k^{\pm} \sin\phi_{\bf k},\tilde{\xi}_{k}^{\pm})$ where $\tilde{\Delta}_k^{\pm}=(\psi_k \mp \gamma_k F_k)$, the single particle energies are $\tilde{\xi}_{k}^{\pm}=\epsilon_{k} \pm \gamma_k \vert G_{\bf k}\vert$ where $\epsilon_{k}=\hbar^2 k^2/(2m)-\tilde{\mu}$, $\tilde{\mu}$ is the chemical potential, and $G_{\bf k}=g(k_x+ik_y)=g k e^{i\phi_{\bf k}}$ with $g$ is the SOC and $\gamma_k=(\vert G_{\bf k}\vert \epsilon_{k}-F_k \psi_k)$. The eigen-energies are given by $E_k^\pm=({\bf h}_{\bf k}^{\pm}.{\bf h}_{\bf k}^{\pm})^{1/2}=[(\tilde{\xi}_{k}^{\pm})^2+(\tilde{\Delta}_k^{(\pm)})^2]^{1/2}$. 

\begin{figure}[here]
\includegraphics [scale=0.52]{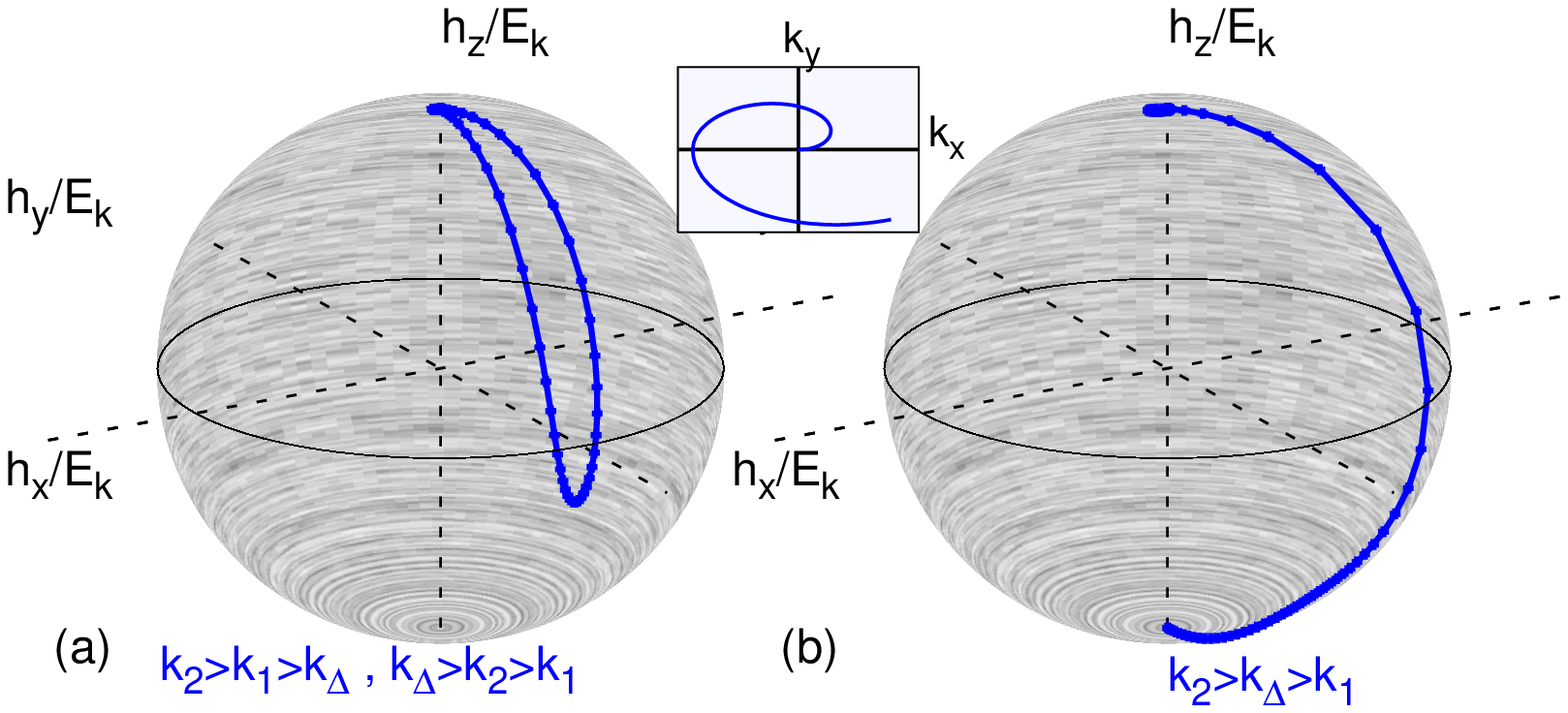}\\
\vskip0.2truecm
\includegraphics [scale=0.33]{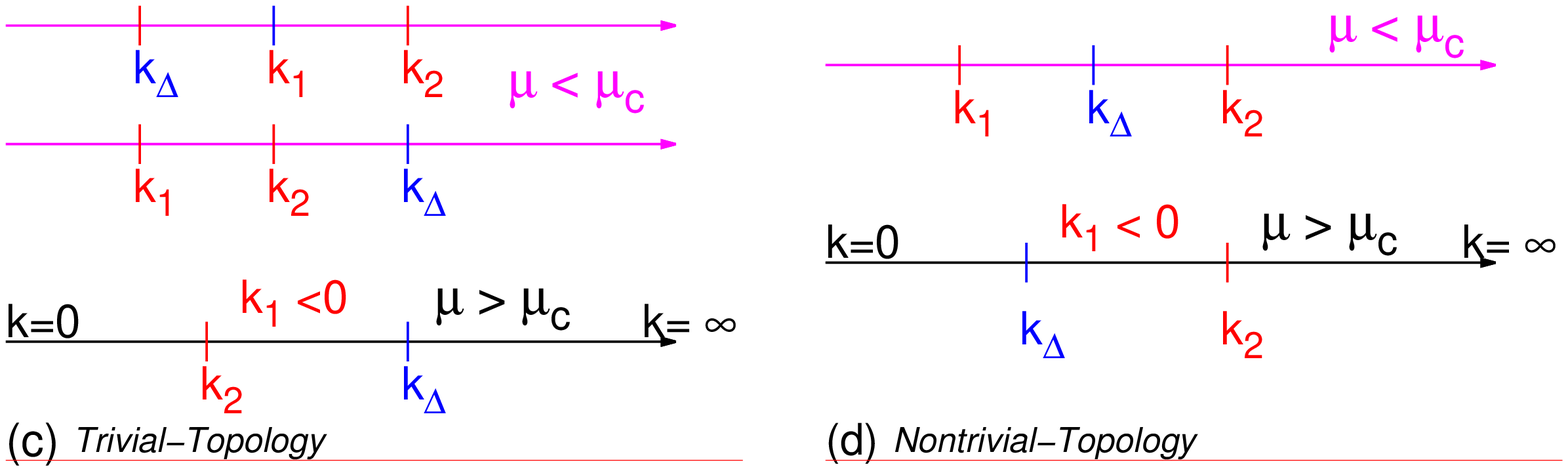}
\vskip-0.2truecm
\caption{(Color online) Nodal positions of $\tilde{\xi}_k^{+}$ and $\Delta_k^{+}$ depicted respectively as $k_1, k_2$ and $k_\Delta$ with different topologies as indicated in a) as trivial, b) as nontrivial. The zeros $k_{1,2}$ are directly determined by the chemical potential and the SOC as $\tilde{\mu}=-\hbar^2k_1k_2/(2m)$ and $g=-\hbar^2(k_1+k_2)/(2m)$. The topology is illustrated on the unit sphere as a) trivial for case (c), and b) nontrivial for case (d).}  
\label{unit_sphere}
\end{figure}

These two branches can have independent topological indices given by the winding of ${\bf h}_{\bf k}^{\pm}$ on $S_2$ and is characterized by the position of the individual nodes in $\tilde{\Delta}_k^{(\pm)}$ relative to the Fermi level. We start the analysis with the {\it kinetic} term $\tilde{\xi}_k^{\pm}=\hbar^2 (\gamma_k k \pm k_1)(\gamma_k k \pm k_2)/(2m)$ where $\pm k_1, \pm k_2$ are the zeros of $\tilde{\xi}_k^{\pm}$. If $\tilde{\xi}_{k_j}^{\pm}=0$ at a positive $k_j, j=1, 2$, then $k_j$ is a Fermi momentum on the $j$'th Fermi surface. We assume without any loss of generality that $ k_2 >  k_1$. For the moment, we take $\gamma_k=1$ and discuss its effect later. The Fermi wavevectors for the $+$ branch are,  
\begin{eqnarray}
k_2=&\frac{m}{\hbar^2}\Bigl[-g+\sqrt{g^2+2\frac{\hbar^2}{m}\tilde{\mu}}\Bigr] \label{energy_nodes_1} \, \qquad &(\tilde{\mu} > 0) \\ 
k_{1,2}=&\frac{m}{\hbar^2}\Bigl[g \mp \sqrt{g^2+2\frac{\hbar^2}{m}\tilde{\mu}}\Bigr] \label{energy_nodes_2}  \, \qquad &(\tilde{\mu} < 0)
\end{eqnarray}
where $\tilde{\mu}=\mu-\mu_c=-\hbar^2 k_1 k_2/(2m)$ and $g=-\hbar^2(k_1+k_2)/(2m)$ are the physical parameters which can be used to shift the Fermi wavevector(s) $k_{1,2}$. All possibilities are shown in Fig.\ref{unit_sphere} for the $+$ branch. The $-$ branch is similarly analyzed.

We will assume that the SOC is sufficiently strong for a large $\vert F_k/\psi_k \vert$ ratio and the gap is allowed to have a node at $k_\Delta$ as $\Delta_k^{(+)}=\delta_+(k)(k-k_\Delta)$ where $\delta_+(k)$ is a smooth function representing the other (irrelevant) details. 
The topology is then determined according to the $Z_2$ classification as shown in Fig.\ref{unit_sphere}.(a,b). In the pure triplet state\cite{pure_triplet_class} when $\psi_k=0$ and $F_k\ne 0$, the topology is controlled by $\mu$ only (i.e for $\mu < \mu_c$  as trivial and  for $\mu > \mu_c$ as non-trivial). In the mixed state however, single parameter characterization is not sufficient due to the additional spin-orbit degree of freedom. The position of $k_\Delta$ in one branch relative to the Fermi wavevector $k_1, k_2$ of the same branch plays important role in the topology.  
 
If $\tilde{\mu}<0$ the kinetic term can have two Fermi wavevectors $k_1, k_2 >0$ given by Eq.(\ref{energy_nodes_2}), or none, i.e. $k_1, k_2 <0$, whereas for $\tilde{\mu}>0$ there is one Fermi momentum, i.e. $k_2$ [as given by Eq.(\ref{energy_nodes_1})]. All possibilities are exploited in Fig.\ref{unit_sphere}.(c, d) for the $+$ branch. The $k_1<k_2<k_\Delta$ configuration is trivial. Using this as a reference, we find that any odd number of Fermi level crossings of $ k_{\Delta}$ switches the $Z_2$ index to the nontrivial topology as shown in Fig.\ref{unit_sphere}.(c, d). 
  
\begin{figure}[H]
\includegraphics[scale=0.35]{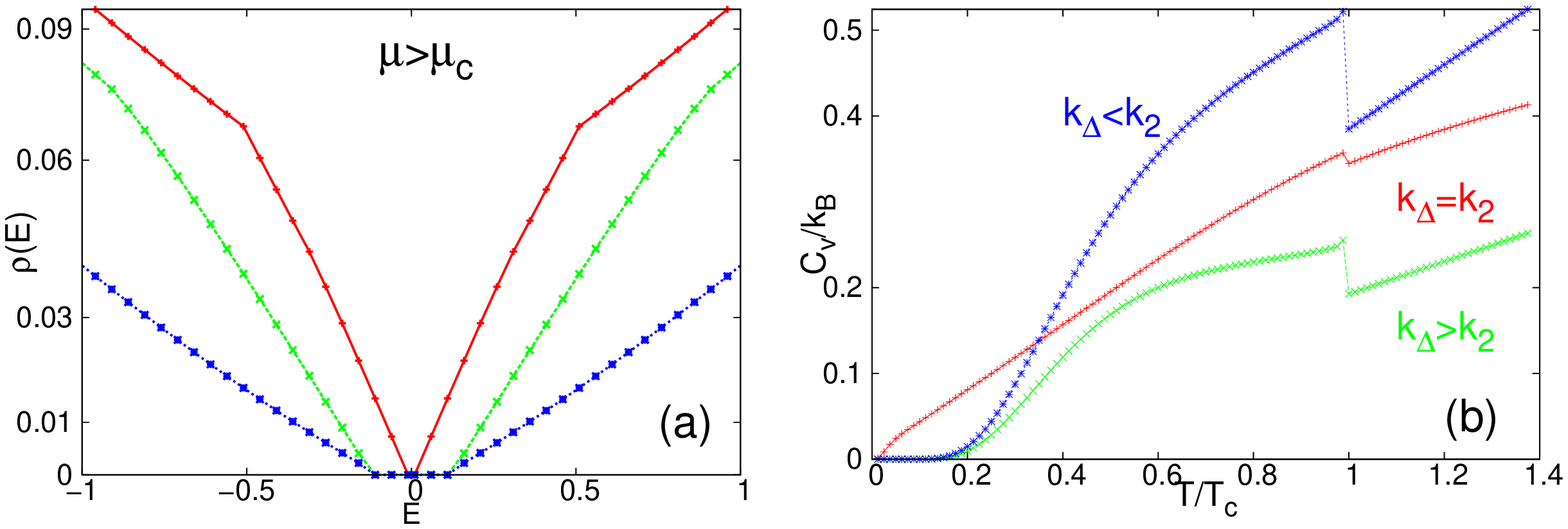} 
\vskip-0.3truecm
  \label{FP}\caption{(Color online) The effect of the Fermi level crossing of the energy gap node $k_\Delta$ for $\mu>\mu_c$ in a) the DOS $\rho(E)$ and b) the $C_V$ corresponding to the cases $k_\Delta < k_1$, $k_\Delta = k_1$ and $k_\Delta > k_1$. Color coding applies to both figures.}
\label{transition_to_from_scaling}
\end{figure}

Experimental observation of the power law behaviour in temperature and the determination of the scaling exponents in NCS have not reached a convincing level especially for weakly anisotropic case. On the other hand, the previous argument indicates that if the RLN positions in Fig.\ref{unit_sphere}.(c,d) can be manipulated, experiments can be performed in a critical range of parameters where the topology changes, which can then have observable thermodynamic behaviour. Thus, RLNs are likely to provide more understanding of the weakly anisotropic conditions. The zeros of the kinetic term $k_1$ and $k_2$ are more sensitive to changes in $\mu$ and $g$ [as shown in Eq.'s(\ref{energy_nodes_1})] whereas, nodes of $\tilde{\Delta}_{k}^{\pm}$ are more sensitive to the interaction potential. It is therefore natural to refer here to the explicit dependence of $k_1$ and $k_2$ on $\mu$ and $g$ which can be used to realize one of the five configurations between $k_1, k_2$ and $k_\Delta$ in Fig.\ref{unit_sphere}.c and d. In Fig.\ref{transition_to_from_scaling} we consider $\tilde{\mu}>0$ with one Fermi momentum at $k_2$ which is tuned externally by $\mu$ and $g$ across $k_\Delta$ in respect of the Eq.(\ref{energy_nodes_1}). The change in the energy DOS, shown in Fig.\ref{transition_to_from_scaling}.a and the calculated behavior of $C_V$ in Fig.\ref{transition_to_from_scaling}.b are distinctively different in the three cases when $k_\Delta$ is below/above the Fermi surface and when it is on it. In the former case, the excitations are absent in the small energy range, hence the $C_V$ is suppressed at small temperatures. However, the suppression is nonexponential, fitting instead into a superposition of power laws $C_V \sim T^\nu$ with exponents depending on $k_\Delta$. In the latter case however, an energy node is created when the gap node is on the Fermi surface. Hence, the energy node lives at the boundary $k_\Delta=k_2$ between two topologically distinct regions as shown in Fig.\ref{unit_sphere}.c and d with $k_2 < k_\Delta$ and $k_2 > k_\Delta$. Arbitrarily low energy excitations are available therefore the DOS and the $C_V$ behave linearly in $T$. 

\begin{figure}[here]
\includegraphics[scale=0.42]{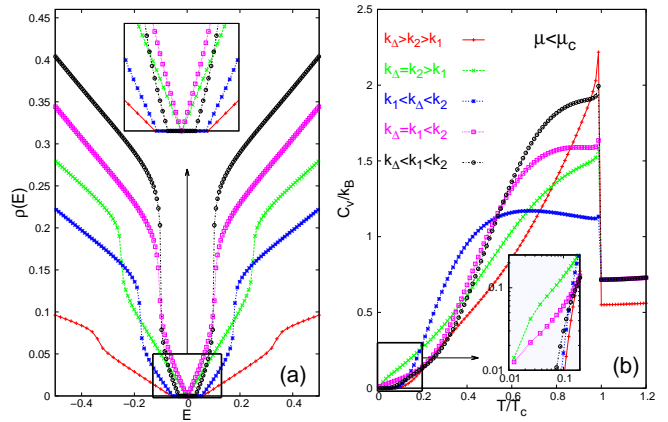}
  \label{FP}\caption{(Color online) The effect of the Fermi level crossing of the energy gap node $k_\Delta$ for $\mu<\mu_c$ on the a) $\rho(E)$ and b) $C_V$ corresponding to 5 different positions of $k_\Delta$ color coded in (b), as also indicated in Fig.\ref{unit_sphere}.(a,b). The insets magnify the low $E$ and low $T$ region of $\rho(E)$ and $C_V$ which are linear for $k_\Delta=k_1$ and $k_\Delta=k_2$.Color coding applies to both figures. 
}
\label{possible_cases}
\end{figure}

f
In the $\tilde{\mu}<0$ case with two different Fermi wavevector, an energy node can occur similarly to the $\tilde{\mu}>0$ case discussed above, when $k_\Delta$ coincides with one of the Fermi wavevectors (i.e. $k_\Delta=k_2$ or $k_\Delta=k_1$), and, according to Fig.\ref{unit_sphere} this is where the topology changes. This is reflected on the DOS and in the power-law dependence of $C_V$ as shown in Fig.\ref{possible_cases} which are frequently reported in experiments for nodal superconductors\cite{Bauer_Sigrist,nodal_thermodynamic}. The specific heat behaves as power-law in regions with distinct topology given by $k_\Delta<k_1<k_2 , k_1<k_2<k_\Delta$ for the trivial as well as $k_1<k_\Delta<k_2$ for the nontrivial cases as depicted in Fig.\ref{unit_sphere}.c and d. The temperature behaviour is a single power law or their admixture  determined by $k_1, k_2$ and $k_\Delta$. The topology changes at each Fermi level crossing when $k_\Delta=k_1$ or $k_\Delta=k_2$ where energy RLNs appear. The $ C_V$ is  linear in temperature as shown in Fig.\ref{possible_cases}.b.  

This argument clearly shows that, RLNs can display a wide range of scaling laws. The linear behavior of $C_V$ in temperature is a manifestation of the existence of high DOS in low energies due to the Dirac-like spectrum in the vicinity of RLNs. An RLN at the Fermi level is more easily identifiable in low temperatures than away from the Fermi level. For instance, in Fig's \ref{transition_to_from_scaling} and \ref{possible_cases} the exponents in the energy gap are much larger than $\nu=2$ that is expected from an ALN leading to a strongly suppressed low temperature behavior. On the other hand the cases $k_\Delta=k_2$ or $k_\Delta=k_1$ in Fig.\ref{transition_to_from_scaling}.b and \ref{possible_cases}.b are easier to identify experimentally but quite rare to find naturally. 

We based our topological analysis on one SOC branch and an important side remark can be made here concerning both branches in the $\tilde{\mu}<0$ case. If $\gamma_k$ changes sign between the two Fermi wavevectors $k_1$ and $k_2$, then both gaps $\Delta_k^{\pm}$ are allowed to have RLNs. This case is interesting but certainly a very rare circumstance. Its experimental identification may be difficult to reveal in thermodynamic measurements, but it may be possible by Andreev reflection spectroscopy-ARS.   

\begin{figure}[h]
\includegraphics[scale=0.4]{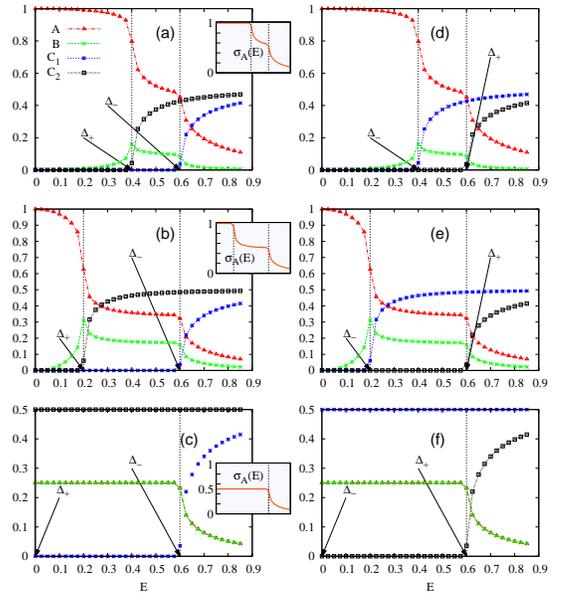}
  \label{FP}\caption{(Color online) Andreev reflection and the transmission coefficients in a semi infinite N-NCS interface are depicted with AC $\sigma_A$ (in units of $2e^2/h$ in the inset for each horizontal case) for three different configurations: a) $\Delta_{\bf k_*}^{-}>\Delta_{\bf k_*}^{+}>0$, b) same as (a) when $\Delta_{\bf k_*}^{+}$ is lowered, and c) same as (a) when $\Delta_{\bf k_*}^{+}=0$. The right column is when $\Delta_{\bf k_*}^{-} \leftrightarrow \Delta_{\bf k_*}^{+}$.}
\label{Andreev_cond}
\vskip-0.5truecm
\end{figure}

The arguments we followed above show that understanding the thermodynamic data in systems with weak anisotropy is obscure and this may be connected with the current confusion in their characterization including the missing physics of RLNs in the whole picture. On the other hand, ARS  and the Andreev conductance (AC) measurements have been useful experimental tools for obtaining information about the pairing symmetry of the s, d and chiral p-wave superconductors.\cite{Bauer_Sigrist} The question here is whether the ARS can probe the RLNs and whether it can be used to separate them from the ALNs. In this context, we calculated the Andreev conductance (AC) at an N-NCS interface where $z<0$ is a normal metal without SOC and $z>0$ is an NCS with SOC. Initially, an electron, spin-polarized in the z-direction, is sent normal to the N-NCS interface from the N side at the energy $E=\hbar^2 k_i^2/(2m)-\mu_N$ where $k_i$ is the wavevector of the incident electron and $\mu_N=\hbar^2 k_F^2/(2m)$ is the chemical potential of the N region. Here $k_F$ is the Fermi wavevector in N side. We also ignore here, the interface potential at $z=0$. The corresponding wavefunctions are \cite{AR_NCS}
\begin{eqnarray}
\Psi_N&=&\{e^{ik_Fz} (1, 0, 0, 0) ^T +a\,e^{ik_Fz} (0, 0, 1, 0)^T \nonumber \\ 
&+&b\,e^{ik_Fz} (0, 0, 0, 1)^T +c\,e^{-ik_Fz} (1, 0, 0, 0)^T \nonumber \\ 
&+& d\,e^{-ik_Fz} (0, 1, 0, 0) ^T \} \qquad {\rm and}
\\
\Psi_S&=&\{ c_1 \,e^{iq_1^+z} ( u_1, \eta \,u_1, v_1, \eta^* \,v_1 ) ^T \nonumber \\ 
&+&c_2 \,e^{iq_2^+z} (-\eta^* \,u_2, u_2, v_2, -\eta^* \,v_2) ^T \nonumber \\ 
&+& c_3 \,e^{-iq_1^-z} (-\eta^* v_1, v_1, u_1, -\eta^* \,u_1)^T \nonumber \\
&+& c_4 \,e^{-iq_2^-z} (v_2, \eta \,v_2, \eta \,u_2, u_2)^T\} \nonumber
\label{wf1} 
\end{eqnarray}
where
\begin{eqnarray}
u_{1 \choose 2}&=&\sqrt{\frac{1}{2}\bigg(1+\sqrt{1-\Big(\frac{\Delta_\pm}{E} \Big)^2}\bigg)} \nonumber\\
\\
v_{1 \choose 2}&=&\sqrt{\frac{1}{2}\bigg(1 - \sqrt{1-\Big(\frac{\Delta_\pm}{E} \Big)^2}\bigg)} \nonumber
\end{eqnarray}
and $\eta=e^{i \phi_{\bf k}}$. Here $a,b$ and $c,d$ ($c_1,\dots,c_4$) are the complex reflection amplitudes for the holes and the electrons (the transmission amplitudes within the NCS in the $\pm$ branches). The specular reflection in the N-region vanishes due to the absence of the interface potential, hence $c=d=0$ and only paralel and antiparalel spin Andreev reflections given by $a,b$ are present. We examine the AC ($ \sigma_A $) between the N and the NCS sides when both of $\Delta^\pm$ are nonzero as shown in Fig.\ref{Andreev_cond}.(a,b,d,e) and, when one of them is zero at a nodal position, i.e. (c,f) therein. The probabilities $A=|a|^2$ $(B=|b|^2)$ are the Andreev reflection probabilities\cite{Bauer_Sigrist,AR_NCS} of the hole in $z<0$ with the same (opposite) spin as the initial electron whereas, $C_{1}=|c_1|^2 (u_1^2-v_1^2)$ ,$ C_2= |c_2|^2 (u_2^2-v_2^2)$ are for the transmission probabilities of the super-current in $z>0$ corresponding to the $\pm$ branches. Here $C_{3,4}=0$ in the NCS due to the semi-infinite geometry. We examine first, the case when both $\Delta^\pm$ are nonzero. In Fig.\ref{Andreev_cond} the Andreev reflection and the transmission coefficients are shown together with the AC calculations (insets of the figures in a,b,c). There, $ \sigma_A $ has a double-step-like behaviour where it starts with a plateau at unit conductance when $E<\Delta^\pm$. If $E$ is between the gaps, the Andreev reflection moves down to a second plateau at $\sigma_A=1/2$ and gradually disappears for higher energies. The second distinct case is when one of the gaps is zero as shown in Fig.\ref{Andreev_cond}.(c,f). There, $\sigma_A$ directly starts at 1/2 at low energies as shown in Fig.\ref{Andreev_cond}(insets of c,f) then going through a single plateau. Hence, the number of nonzero energy gaps is observable in the number of plateaus. 

A clear distinction between RLNs and ALNs in an NCS can thus be made easily in an AC measurement. The former is identified by the plateau structure in the Fig.\ref{Andreev_cond}. On the other hand the Andreev reflection should vanish for an ALN at normal incidence and should be anisotropic at specular angles of incidence\cite{d-wave}.        

In summary, theories incorporating exact momentum dependence of the OPs\cite{TH_MG} indicate the existence of the RLNs which should be incorporated into our knowledge of the NCSs for the complete treatment of the nodal dynamics. We demonstrated that the RLNs can be more pronounced in systems with weak anisotropy and they can be effective in the low temperature anomalies that are not explained by the ALNs. The topology around the RLNs which is classified by the relative position of the gap node with respect to the Fermi wavevector, is determined by the chemical potential and the SOC. Finally, the Andreev spectroscopy is an efficient tool for probing the RLNs in N-NCS junctions which, with its specific conductance plateaus, shows clear distinction from the better known ALNs.


\begin{thebibliography}{99} 
\bibitem{He3}R. Balian and R.N. Werthamer {\bf 131}, 1553 (1963); P.W. Anderson and P. Morel {\bf 123}, 1911 (1961); A.J. Leggett, Phys. Rev. Lett. {\bf 29}, 1227 (1972).

\bibitem{Heavy_Fermion}  R. H.  Heffner  and  M. R.  Norman,  Comments  Condens. Matter {\bf 17}, 361 (1996);H.R. Ott, H. Rudigier, Z. Fisk and J.L. Smith, Phys. Rev. Lett. {\bf 50}, 1595 (1983);F. Steglich, J. Aarts, C.D. Bredl, W. Lieke, D. Meschede, W. Franz and H. Schafer, Phys. Rev. Lett.{\bf 43} , 1892 (1979).

\bibitem{High_Tc}J.G. Bednorz and K.A. M\"{u}ller, Z. Phys. {\bf B64}, 189 (1986); Patrick A. Lee, Naoto Nagaosa, and Xiao-Gang Wen Rev. Mod. Phys. {\bf 78}, 17 (2006). 

\bibitem{TH}T. Hakio\u{g}lu and M. \c{S}ahin, Phys. Rev. Lett. {\bf 98}, 166405 (2007); M.A. Can and T. Hakio\u{g}lu, Phys. Rev. Lett. {\bf 103}, 086404 (2009).

\bibitem{TRSB}Wei Liu and Alexander Punnoose, Phys. Rev. Lett., {\bf 114}, 176402 (2015),Luke, G. M., Y. Fudamoto, K. M. Kojima, M. I. Larkin, J. Merrin, B. Nachumi, Y. J. Uemura, Y. Maeno, Z. Q. Mao, Y. Mori, H. Nakamura, and M. Sigrist,  Nature (London) {\bf 394}, 558 (1998).

\bibitem{OPs} K. Izawa, H. Takahashi, H. Yamaguchi, Yuji Matsuda, M. Suzuki, T. Sasaki, T. Fukase, Y. Yoshida, R. Settai, and Y. Onuki ,Phys. Rev. Lett. {\bf 86}, 2653 (2001) ;P. A. Frigeri, D. F. Agterberg, A. Koga, and M. Sigrist Phys. Rev. Lett. {\bf 92}, 097001 (2004).

\bibitem{evidence1} H. Q. Yuan, D. F. Agterberg, N. Hayashi, P. Badica, D. Vandervelde, K. Togano, M. Sigrist, and M. B. Salamon Phys. Rev. Lett.{\bf 97}, 017006(2006); Y. Maeno, S. Kittaka, T. Nomura, S. Yonezawa, and K. Ishida, J. Phys. Soc. Jpn. {\bf 81}, 011009 (2012); K. Ishida, H. Mukuda, Y. Kitaoka, K. Asayama, Z. Q. Mao, Y. Mori, and Y. Maeno, Nature (London) {\bf 396}, 658 (1998); Y. Maeno, H. Hashimoto, K. Yoshida, S. Nishizaki, T. Fujita, J. G. Bednorz, F. Lichtenberg, Nature (London) {\bf 372}, 532 (1994); A.P. Mackenzie and Y. Maeno, Rev. Mod. Phys. {\bf 75}, 657 (2003).
\bibitem{evidence2}  K. Izawa, Y. Kasahara, Y. Matsuda, K. Behnia, T. Yasuda, R. Settai, and Y. Onuki Phys. Rev. Lett. {\bf 94}, 197002(2005); Ismardo Bonalde, Werner Brämer-Escamilla, and Ernst Bauer, Phys. Rev. Lett. {\bf 94}, 207002 (2005);J. Chen, M. B. Salamon, S. Akutagawa, J. Akimitsu, J. Singleton, J. L. Zhang, L. Jiao, and H. Q. Yuan,Phys. Rev. {\bf B 83}, 144529 (2011). 
\bibitem{Sigrist_Ueda} M. Sigrist and K. Ueda, Rev. Mod. Phys. {\bf 63}, 239 (1991); A. B. Vorontsov, I. Vekhter and M. Eschrig, J. Phys. Soc. Jpn. {\bf 77}, Suppl. A: 165 (2008).
\bibitem{Weyl} Tobias Meng and Leon Balents Phys. Rev. {\bf B 86}, 054504(2012).
\bibitem{TH_MG} "{\it Unconventional Pairing Symmetries and Double Line Nodes From a Single Model}" T. Hakio\u{g}lu and M. G\"{u}nay, arXiv:1411.4273 (2014).
\bibitem{ARPES} C. N. Veenstra, Z.-H. Zhu, M. Raichle, B. M. Ludbrook, A. Nicolaou, B. Slomski, G. Landolt, S. Kittaka, Y. Maeno, J. H. Dil, I. S. Elfimov, M. W. Haverkort, and A. Damascelli,Phys. Rev. Lett. {\bf 112}, 127002 (2014).
\bibitem{Andreev_cond} Mintu Mondal, Bhanu Joshi, Sanjeev Kumar, Anand Kamlapure, Somesh Chandra Ganguli, Arumugam Thamizhavel, Sudhansu S. Mandal, Srinivasan Ramakrishnan, and Pratap Raychaudhuri, Phys. Rev. {\bf B 86}, 094520 (2012).
\bibitem{Kurter} A. P. Zhuravel, B. G. Ghamsari, C. Kurter, P. Jung, S. Remillard, J. Abrahams, A.V. Lukashenko, A. V. Ustinov and S. M. Anlage, Phys. Rev. Lett. {\bf 110}, 087002 (2013). 
\bibitem{Bauer_Sigrist} {\it Non-Centrosymmetric Superconductors: Introduction and Overview (Lecture Notes in Physics)}", Ed's E. Bauer and M. Sigrist, Springer, (2012).
\bibitem{nodal_thermodynamic} B. Mazidian, J. Quintanilla, A. D. Hillier, and J. F. Annett Phys. Rev. {\bf B 88}, 224504 (2013). 
\bibitem{strong_anisotropy} N. Momoto and M. Ido, Physica (Amsterdam) {\bf 264C}, 311 (1996); T. Sakon et al., Physica (Amsterdam){\bf 199C}, 154 (1994); R. Movshovich, M. Jaime, J. D. Thompson, C. Petrovic, Z. Fisk, P. G. Pagliuso, and J. L. Sarrao, Phys. Rev. Lett. {\bf 86}, 5152 (2001).
\bibitem{weak_anisotropy} H. Takeya et al. , Phys. Rev. {\bf B 72} , 104506 (2005); G. Eguchi, D. C. Peets, M. Kriener, S. Yonezawa, G. Bao, S. Harada, Y. Inada, G.-q. Zheng, and Y. Maeno Phys. Rev. {\bf B 87}, 161203(R) (2013);
\bibitem{inadequate}I. Bonalde, H. Kim, R. Prozorov, C. Rojas, P. Rogl, and E. Bauer, Phys. Rev. {\bf B 84}, 134506 (2011) ; P. Hafliger, R. Khasanov, R. Lortz, A. Petrovic, K. Togano, C. Baines, B. Graneli, and H. Keller, J. Supercond. Novel Magn. {\bf 22}, 337 (2009)
\bibitem{pure_triplet_class} Y. Maeno, S. Kittaka, T. Nomura, S. Yonezawa and K. Ishida, J. Phys. Soc. Japan, {\bf 81}, 011009 (2012).
\bibitem{AR_NCS} Yokoyama, T., Tanaka, Y., Inoue, J. Phys. Rev.{\bf  B 72}, 220504(R) (2005).
\bibitem{d-wave} Yukio Tanaka and Satoshi Kashiwaya, Phys. Rev. Lett. {\bf 74}, 3451(1995).
\end{thebibliography}
\end{document}